\documentclass[runningheads]{llncs}

\usepackage{enumitem}
\usepackage{multirow}
\usepackage{graphicx}
\usepackage{balance}
\usepackage[caption=false]{subfig}
\usepackage{xcolor}
\usepackage[misc]{ifsym}

\begin{document}

\title{Experts' View on Challenges and Needs for Fairness in Artificial Intelligence for Education}

\titlerunning{Experts' View on Fairness in Artificial Intelligence for Education}

\author{Gianni Fenu \orcidID{0000-0003-4668-2476} \and
Roberta Galici \orcidID{0000-0002-1492-3514} \and
Mirko Marras \orcidID{0000-0003-1989-6057}
}

\authorrunning{G. Fenu et al.}

\institute{University of Cagliari, Cagliari, Italy \\ 
\email{\{fenu, roberta.galici\}@unica.it, mirko.marras@acm.org} 
}

\maketitle             

\begin{abstract}
In recent years, there has been a stimulating discussion on how artificial intelligence (AI) can support the science and engineering of intelligent educational applications. Many studies in the field are proposing actionable data mining pipelines and machine-learning models driven by learning-related data. The potential of these pipelines and models to amplify unfairness for certain categories of students is however receiving increasing attention. If AI applications are to have a positive impact on education, it is crucial that their design considers fairness at every step. Through anonymous surveys and interviews with experts (researchers and practitioners) who have published their research at top-tier educational conferences in the last year, we conducted the first expert-driven systematic investigation on the challenges and needs for addressing fairness throughout the development of educational systems based on AI. We identified common and diverging views about the challenges and the needs faced by educational technologies experts in practice, that lead the community to have a clear understanding on the main questions raising doubts in this topic. Based on these findings, we highlighted directions that will facilitate the ongoing research towards fairer AI for education. 
\end{abstract}

\section{Introduction}
Educational systems that rely on artificial intelligence (AI) are increasingly influencing \emph{the quality of the education we receive}. Notable examples of AI-based models integrated so far include early predictors of student success (e.g., \cite{rastrollo2020analyzing}), clustering techniques for learner modelling (e.g., \cite{abyaa2019learner}), intelligent tutoring and scaffolding (e.g., \cite{abidi2019prediction}), agents for motivational diagnosis and feedback (e.g., \cite{alsuliman2020machine}), and models for recommending peers or learning material (e.g., \cite{wang2020expert}). With this growth, the potential of AI to \emph{amplify unfairness in educational applications} has received growing attention in both academia and industry as well as the press. Indeed, articles in mainstream media have reported systemic unfair behaviors of some AI-enabled educational systems. For example, an automated college enrolment system more likely to recommend enrolments from certain ethnic, gender, or age groups \cite{britto2019machine}, or a machine-learning system for evaluating PhD applicants in Computer Science that exacerbates current inequality in the field \cite{shahbazi2020toward}.

Concerted effort in the area of fairness in educational applications has mainly focused on the \emph{design and operationalization of fairness definitions} \cite{verma2018fairness} and \emph{algorithmic methods to assess and mitigate undesirable biases} in relation to these definitions \cite{hajian2016algorithmic}. Other works have studied fairness in educational AI systems through a social and psychological lens \cite{berendt2020ai}. As the field matures, literature reviews are collecting definitions, methods, and perspectives in a unified framework \cite{holstein2019improving}. Specialized initiatives, such as workshops \cite{alameda2020fate} and special issues \cite{data7020021}, are also focusing on biases and unfairness in educational AI. While some fair AI studies are already being researched, they often represent isolated examples. If the resulting applications are to have a positive impact on education, however, \emph{fairness-aware practices should become common for any person while creating an educational application that leverages AI}. Being informed about the actual challenges and needs for supporting the development of \emph{fairer AI for education} is hence crucial.

Creating AI-based educational systems raises many unique challenges not commonly faced with intelligent systems in other domains \cite{renz2020prerequisites}. In the broader AI field, several papers have dealt with fairness \cite{caton2020fairness,iosifidis2019fae,mitchell2018prediction,mehrabi2021survey}. Despite this attention to unfairness, to the best of our knowledge, only two prior studies have investigated challenges and needs for supporting the creation of fairer AI by directly interrogating experts \cite{holstein2019improving,veale2018fairness}. Unlike those studies, focused on the public-sector and commercial AI practitioners across a range of high-stakes contexts, our study focuses on educational researchers and practitioners (referred for convenience as experts) including AI in their workflow, who usually have experience in developing AI systems, but are relatively new to thinking about fairness. Considerations, beliefs, practices, motivations, and priorities in integrating fairness may be less clear within these contexts, applications, and cultures. 

Our study in this paper investigates the challenges and needs faced by the educational community, whose products are going to affect the education of individuals, in \emph{integrating and monitoring for unfairness and taking appropriate action}. Through an anonymous survey of 136 educational researchers and  practitioners who have published their research at top-tier educational conferences in 2021 (e.g., AIED, EDM, and L@S), we analyze teams’ existing opinions and experience around the development of fair educational AI, as well as their challenges and needs for support. To deeper the key themes of our survey, we also conducted semi-structured interviews with 28 of those researchers and practitioners. To our knowledge, this is the \emph{first systematic investigation of experts’ challenges and needs around fairness in educational AI}. 

Through our investigation, we identify a range of real-world needs often not stated in the literature so far, as well as several common areas. For example differently from the broader AI field, large-scale data collection is not always considered as a solution, since biases are driven by complex reasons to be understood in the local context.  Such concerns are also extended to research teams’ own blind spots, since teams often struggled to anticipate the sub-populations and forms of unfairness they need to consider for specific kinds of applications. Moreover, though fair educational AI has overwhelmingly focused on data collection issues, assessment and debiasing of unfairness are also crucial, e.g., by having continuous fairness assessment at all stages of the pipeline. Because fairness can be context and application dependent, domain-specific educational resources, metrics, processes, and tools are urgently needed, such as to open data and source code for public scrutiny and create participatory processes for fairness checking. Another identified area is the development of auditing processes and tools, to make fairness issues emerge. Based on these findings, we highlight opportunities to have a greater impact on the fair educational AI landscape.

\section{Methodology}
\begin{figure}[!t]
\centering
  \includegraphics[width=0.95\linewidth]{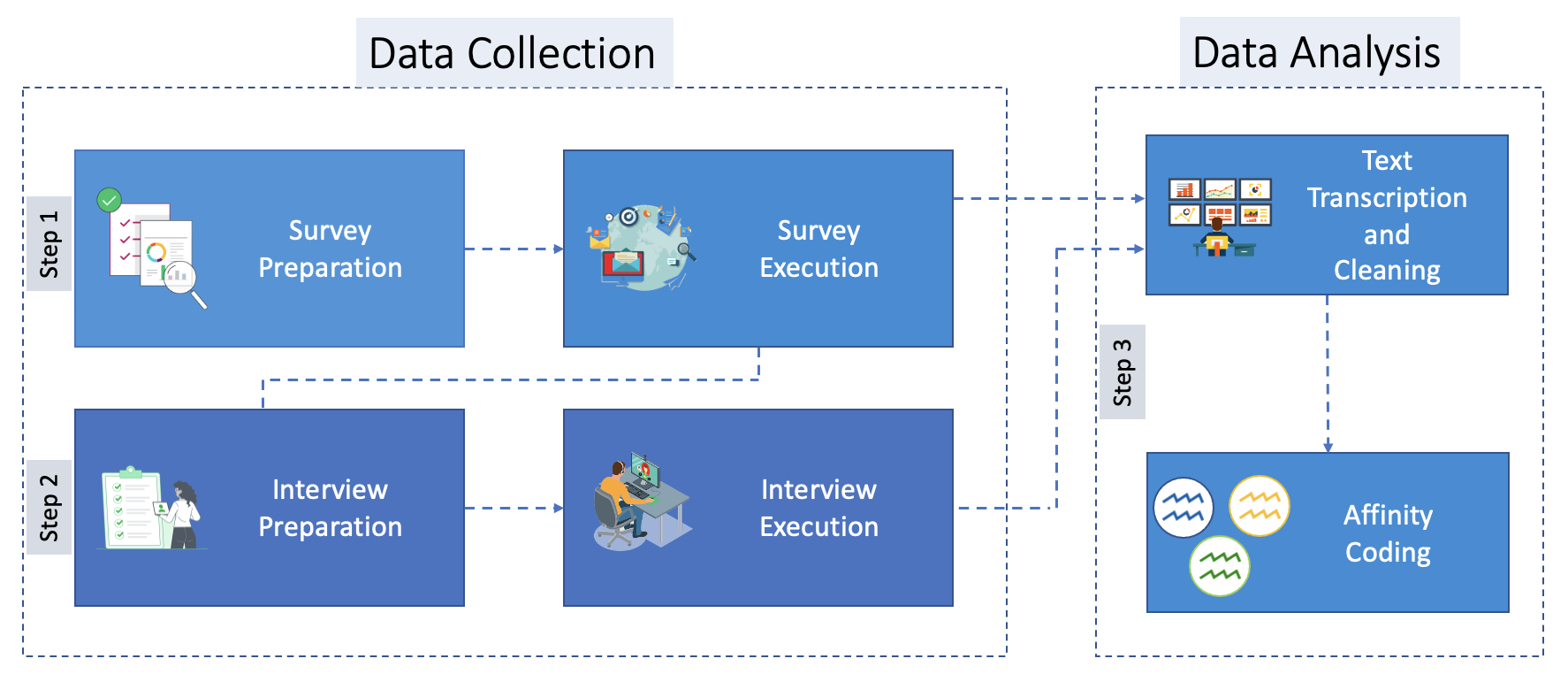}
  \vspace{-5mm}
  \caption{We first conducted an anonymous survey (Step 1). To deeper the key themes from the survey, we then conducted semi-structured, one-on-one interviews (Step 2). We finally analyzed answers from both the surveys and the interviews (Step 3).}
\vspace{-3mm}
  \label{fig:method}
\end{figure}

The goal of this study is to investigate the challenges and needs faced by the educational community, whose products are going to affect the education of individuals, in integrating and monitoring for unfairness and taking appropriate action. To this end, we adopted a two-step mixed approach, depicted in Fig. \ref{fig:method}. 

\subsection{Survey Study Implementation}
In a first step, to get a broad sense of challenges and needs for addressing fairness while developing educational AI, we conducted an anonymous online survey. 

\begin{figure*}[t]
\centering
\subfloat[Survey\label{fig:survey}]{
    \includegraphics[width=0.5\linewidth]{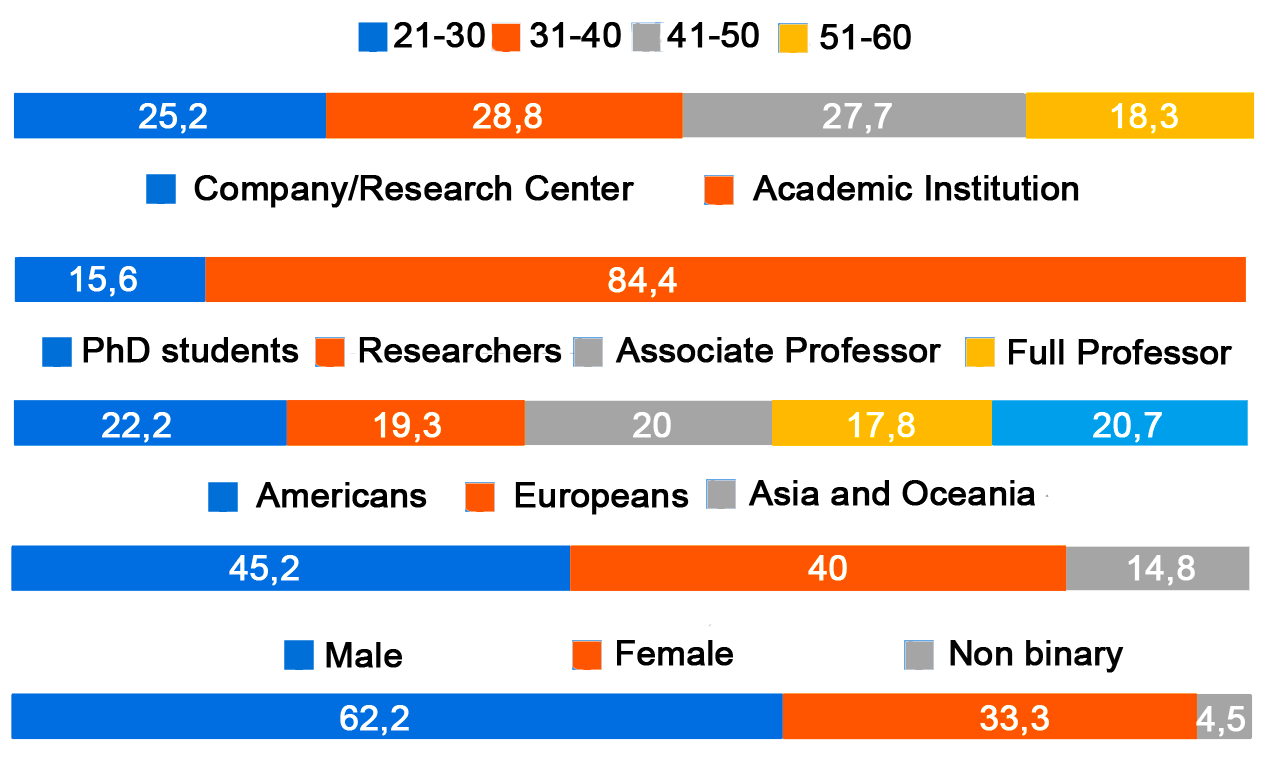}}
\subfloat[Interview\label{fig:interview}]{
    \includegraphics[width=0.5\linewidth]{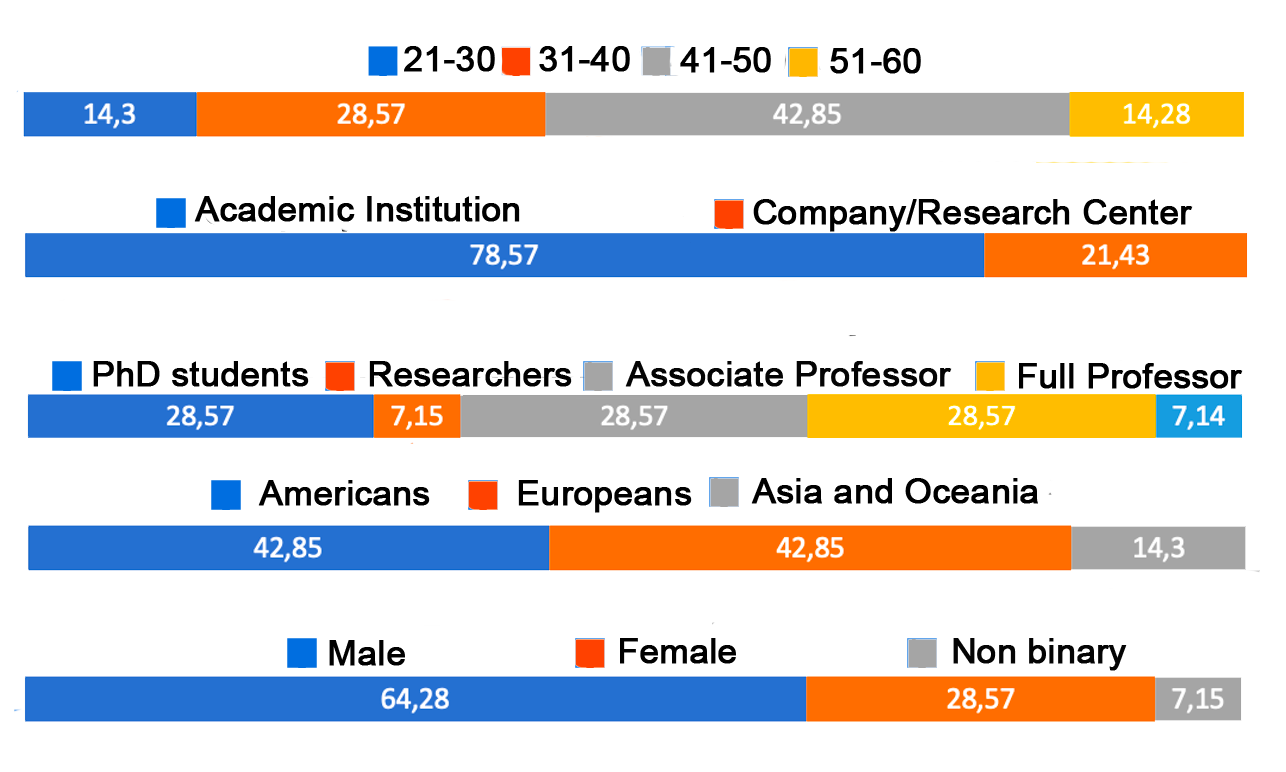}}
\vspace{-2mm}
\caption{Sample population statistics for our survey and interview process.}
\vspace{-3mm}
\label{fig:distribution}
\end{figure*}

\vspace{2mm} \noindent \textbf{Participants Recruitment and Statistics}. The survey participants were recruited using a systematic process, to make sure that our study was not based on opinions of an arbitrarily selected population. Specifically, we systematically scanned the 2021's proceedings of the top-tier educational conferences in a manual process, namely AIED, EAAI, EC-TEL, EDM, ICALT, ITS, LAK, and L@S, to identify the authors who had a paper accepted. We also considered the authors of papers in the special issues about fair educational AI in IJAIED. We then directly emailed the survey to those authors between Sep and Dec 2021, and often invited them to pass the survey on to colleagues working on educational AI, within their organization. In total, we contacted 2,175 experts, and 136 out of them (6\%) completed at least one section beyond demographics. A description of the population who answered to the survey is provided in Fig. \ref{fig:distribution}a.

\vspace{2mm} \noindent \textbf{Survey Execution}. We structured the survey as a Google Form and developed survey questions to investigate the prevalence and generality of emerging themes. First, we asked a set of demographic questions to understand our respondents’ provenience and backgrounds, including their technology area(s) and role(s). In a branching sequence of survey sections, respondents were then asked about their opinions, challenges, and needs for support around fairness, with each section pertaining to one stage of the educational AI development pipeline\footnote{A pdf copy of the survey questions is available at \url{https://bit.ly/FairAIEdSurvey}.}. For each of the latter questions, we provided open-ended response options that allowed respondents to elaborate on their arguments. We finally asked them to leave their email address in case there were willing to participate in a subsequent interview.  

\subsection{Interview Study Implementation}
In a second step, to validate and deeper our findings from the previous survey answers, we then conducted a series of semi-structured, one-to-one interviews. 

\vspace{2mm} \noindent \textbf{Participants Recruitment and Statistics}. In Jan 2022, we involved the experts who answered the survey and gave their availability to participate in a follow-up interview by providing their e-mail addresses in the last question of the survey. In total, 29 out of 136 survey respondents (21\%) were willing to participate in this second step, across as many research teams. Whenever possible, we tried to interview people in different roles on the same team to hear (potentially) different perspectives. Each interview lasted 30 minutes and was conducted remotely since involved people came from a diverse set of countries. Key statistics about our sample sub-population are depicted in Fig. \ref{fig:distribution}b. 

\vspace{2mm} \noindent \textbf{Interview Execution}. Each participant was first reminded of the purpose of our research. Then, the interview focused on the participant awareness about the debate and research on demographic fairness in educational AI and the most important challenges and open questions in the area in general (questions $7$ and $8$ of the survey).  Each participant was then asked to better clarify the educational AI applications they are working on and who the target users of these applications are (question $8$ of the survey). Interviewees were then asked whether fairness is regularly considered in their workflow and what it means for them to be fair in their applicative context (question $9$ of the survey). Discussion points were often prompted from the survey answers.

We then provide survey questions about fairness at each stage in their educational AI development pipeline, from collecting data to designing datasets to assessing and potentially mitigating fairness issues (questions $10$ to $13$ of the survey). For each of these stages, interviewees were asked a broad opening question in line with the one reported in the corresponding question of the survey and follow-up questions based on the answers provided in the survey. This follow-up led interviewees to reflect more deeply on their practices. 

\subsection{Survey and Interview Data Analysis} 
We integrate the answers collected from each survey question with the corresponding interview counterpart. Interviews and surveys were identified with an ID, and the same ID was used if both come from the same participant. To synthesize findings using standard methodology from contextual design, we conducted interpretation sessions and adopted affinity diagramming (e.g., see also \cite{liu2021adqda}). Specifically, we employed bottom-up affinity diagramming to iteratively generate codes for various individual texts and then group these codes into successively higher-level themes. The themes emerged from the data rather than being imposed. Key themes are presented in the following section. 

\section{Results and Discussion}
In the following, we discuss current challenges and needs around fairness, organized by top-level themes associated to the survey questions and the interview-based deepening (the latter served as a confirmation and enrichment to the survey responses), granularly framed according to the resulting affinity diagram. These include needs and challenges on data collection and modelling (questions $9-10$), unfairness detection and mitigation (question $11$), fairness guarantees provision (question $12$), holistic fairness auditing (question $13$), paired with systemic aspects such as team composition, cross-organization collaborations, and educational AI maturity (question $7$). Within each top-level theme, we present selected sub-themes. \emph{It should be noted that our study aims more to uncover open questions than providing answers. The latter requires further discussion and work of the research community as as whole}, also driven by our study. 

\subsection{Challenges and needs in data collection and grouping (Q9-10)}

\vspace{2mm} \noindent \textbf{Cultural dependencies in demographic representation}. The majority of the participants recognizes that researchers, which are often not demographically representative of their societies, tend to involve people at their hand (their students), which are often not representative either.
For instance, in question $10$ of the survey, a participant reported that \emph{"it is difficult to collect data enough representative of different contexts, such as countries, universities, and society, due to different culture, viewpoints and rules"}. The same participant, during the interview, further observed that \emph{"most research focuses on certain countries, and is done in English, so findings are more representative of certain societies and educational systems"}. Overall, it was often pointed out that no dataset represents the diversity of the population, leaving always some people underrepresented.

\vspace{2mm} \noindent \textbf{Biases are driven by reasons to be understood in the local context}. Paired with the above point, differently from the broader AI field, several participants do not think that large-scale data collection will really get at the nuances of fairness in educational AI. For instance, in question $10$ of the survey, a participant highlighted that \emph{"unlike some AI applications where very broad groups are relevant (e.g., face recognition), biases in education are driven by complex reasons to be understood locally"}. This clearly calls for localized data collection paired with data sharing practices. 

\vspace{2mm} \noindent \textbf{Hidden relationships between demographics and learning variables}. Our participants found that it is generally difficult to identify issues that really drive fairness. For example, during the interview, a participant reported that \emph{"in some cases, ethnicity does not directly cause differences in how students interact with educational software or the data coming out of it, but rather students' life experiences that are correlated with ethnicity"} (e.g., feeling uncomfortable in class because discrimination). In general, it was observed that different demographic groups might have different ways of responding to psychological measures. A participant envisioned during the interview that \emph{"educational AI models might need to be demographically stratified"}. Overall, challenges emerged on what the demographic attributes really proxy for and how experts can measure that. 

\vspace{2mm} \noindent \textbf{Giving individuals continuous control of their data}. A vast segment of our participants acknowledges that individuals should always have complete access and control over their data as well as new data created about them. One participant, following up on what reported in the survey answer, suggested that \emph{"access should be controlled in such a way that confidential information will not be inadvertently shared beyond their control"}. From our affinity diagram, it seems clear the need of supporting tools to inform the users about which data the system is using and for what. In this sense, another participant envisioned in the survey that \emph{"in these tools, the user may turn on and off on which data they think the system needs to use"}. Overall, challenges and needs emerged on letting individuals have  control on which of their data is used in any educational system. 

\subsection{Challenges and needs of fairness-aware technical pipelines (Q11)}

\vspace{2mm} \noindent \textbf{Continuous fairness assessment at all stages of the pipeline}. Most of our participants emphasized that fairness should be taken into account from the beginning, and that all choices (data, optimization criteria, interventions) should be viewed from a fairness point of view. For instance, in question $11$ of the survey, a participant envisioned that \emph{"aims and objectives of the data collection have to be clearly defined and negotiated with the participants, through explicit discussions"}. Fairness is recognized as to be part of the design process from multiple lenses, from developing a team which can reach a high level of expertise in fairness until deploying the system. Overall, protocols and guidelines on how to include fairness through the pipeline are therefore needed. 

\vspace{2mm} \noindent \textbf{Understanding and acknowledging weaknesses of the system}. In designing educational AI systems, a segment of our participants recognizes as important to understand where systems work and the cases where they may suffer. In the survey, a participant suggested that \emph{"this begins by understanding the scope and limitations of the data on which the systems are based, since it is often infeasible to achieve full transparency or explainability in regard to these systems"}. The way of working could also enable to keep the models as they are but inform users about the difference in accuracy between certain demographic groups, for instance. Overall, there is a need to acknowledge these aspects continuously to reach a better understanding of the strengths and limitations of the systems.

\vspace{2mm} \noindent \textbf{Reducing frictions between model effectiveness and fairness}. Our participants consistently mention that the main challenge is how to balance the accuracy of predictions with fairness. For example, a participant mentioned in the survey that \emph{"we might get a good accuracy while directly using demographic features (e.g., gender), but the value of those features might be that they encode something else (e.g., prior experience)"}. This leads for instance to observe that models can achieve the same performance without using demographic features (e.g., prior experience). Moreover, if a model performs really well for one group and poorly for another, a participant found \emph{"debatable whether the benefit should be withheld from the group it works well for"}. There may be alternative practices that still lead to fair uses, even if fairness is not incorporated into a model.

\vspace{2mm} \noindent \textbf{Creating cross-institutional frameworks for addressing fairness}. Some respondents highlighted the need of creating a consortium of organizations (e.g., universities and companies) from different countries and defining a unified framework for data collection and fairness-aware model evaluation applicable in all those universities. For instance, a participant in the interview said that \emph{"there should be an increasing trust between government, institutions, researchers, and practitioners to access sensitive data, building on top of privacy regulations for de-identified data sharing in educational systems"} (e.g., by accessing data at institution level). However, some participants identified that leveraging data, even when anonymized, to improve educational systems is still challenging.  

\subsection{Challenges and needs in providing fairness guarantees (Q12)}

\vspace{2mm} \noindent \textbf{Opening data and source code for public scrutiny}. In this line, a range of participants highlighted that people who develop educational AI systems should publish or release models and analyses for public scrutiny, especially when there are concerns about their models being unfair. For instance, a participant found important that \emph{"sharing data, source code, and pre-trained models in open online repositories should be encouraged"}. Overall, this practice clearly goes in contrast to copyright, therefore creating guidelines and directives that regulate this sharing process is a major requirement to advance from these perspective.  

\vspace{2mm} \noindent \textbf{Fairness should not be a property of the model only}. In response to question $13$ of the survey, several participants emphasized that a common practice is to try and offer as fair results as possible in AI models, but there might still be biases when using these models as a final service to the user (i.e., the way the model is deployed in the complex educational ecosystem). For instance, a participant reported that \emph{"fairness should be therefore a property of the offered service as well at the end"}. Hence, there is a need, for a large segment, to envision guidelines and practices that embed fairness as a constraint or metric for the underlying predictive model as well as a key indicator for the service.

\vspace{2mm} \noindent \textbf{Showing explicit evidence the system's potential unfair impacts}. A segment of the participants believes that institutions adopting educational AI need data that supports a claim of the tool being fair. A participant believes that \emph{"such aspect should not be explainable or transparent to students, since it is probably the case that drawing students' attention to demographics invokes stereotype threat, which might be in contrast with other participants ideas"}. For another segment, students should have the right to know how a system works and be informed on any shortcoming about fairness that the system might have. \emph{"They should not feel that a system might guide them to take a decision because some fairness guarantees of the system are not met by the system"} is an example reported by another participant. Another interesting aspect identified by our participants is that the extent of transparency, accountability, and explainability should depend on the level of impact the system has. For example, a participant reported that \emph{"decisions on weather a student must repeat a course should probably offer more accountability than small recommendations on a platform"}.

\vspace{2mm} \noindent \textbf{Creating participatory processes for fairness checking}. A segment of our participants identified that, ideally, there should be an independent third-party entity that should be able to provide sample data to the educational AI service and then assess whether such service is fair. During the interview, a participant followed up on this point, envisioning that \emph{"all aspects should be showcased to an ethical commission within the organization or that a learner advocate should be allowed to conduct exploratory research into how the system might have detrimental effects on later learner success"} (e.g., showing the system is fair by finding and correcting negative consequences). Who and how should be involved is an open question for the community and is expected to be defined by future advances. 

\vspace{2mm} \noindent \textbf{Regulations for defining responsibilities around fairness issues}. Several participants found that, in fair educational AI systems, it is important to define who is the guarantor of the system's fairness and what are the consequences for not living up to the guarantee. As an example, a participant often compared this aspect to \emph{"service level agreements provided by cloud services, where failing to live up to a guarantee might result in a financial penalty"}. Several of the participants emphasized that it should become mandatory to guarantee an overall fair treatment and to certify that the model is fair according to certain variables.

\subsection{Challenges and needs of a more holistic fairness auditing (Q13)}

\vspace{2mm} \noindent \textbf{Human-centered evaluation of fairness}. Most of the participants stated that the evaluation should be human-centered and cover different levels of analysis. A participant, while deepening question $13$ of the survey during the interview, \emph{"the evaluation should consider statistical metrics, expert audits regarding system design and training data sets, and meetings with stakeholders representing the most impacted groups"}. From the responses, it emerged that this multi-level evaluation gives the opportunity to make sure that the evaluation protocol is properly adapted to the specific applicative case. To make the protocol more efficient, some respondents also suggested that some parts of the evaluation protocol can be automated, but stakeholders must stay involved in any case.

\vspace{2mm} \noindent \textbf{Creation of tools that allow stakeholders to audit models}. Our participants highlighted making educational AI transparent to the end users is essential and envision a need of letting stakeholders analyze system data for fairness and outcomes, of course. A participant dug deeper into this aspect during the interview, saying that \emph{"it would be necessary to have experts doing this, otherwise other stakeholders will have to spend a lot of time and resources"}. Overall, a common view is that the student (or the instructor) should understand why they are given a certain prediction so that they can reflect and react constructively. Some of the respondents observed that it might be tough to deeply explain AI-driven educational predictions to students, especially young students, though they should always have some high-level understanding of what the system is doing, also after some specific training on this task.

\vspace{2mm} \noindent \textbf{Contextualized and application-specific properties to inspect}. The majority of the participants raised doubts regarding the extent to which the fairness metrics and protocols defined in the broader AI community can also work for educational AI systems. For instance, a participant expected that \emph{"the fairness spectrum for AI in the educational field should be investigated, and a framework to be adapted to every context and application might be the outcome"}. Overall, tailored frameworks of processes and properties (also depending on the local data privacy and protection laws) should be better aligned to the specificity of the educational field, rather than being merely built upon those of other domains as black boxes. Indeed, education is recognized as a highly human-centered rather than data-centric field, and tailored frameworks become almost mandatory. 

\vspace{2mm} \noindent \textbf{Long-term learning-related evaluation of fairness}. Our participants generally emphasized that it exists an overly computational definition of fairness in the broader AI field, that tends to evaluate the demographic differentials instead of identifying the strengths and weaknesses of each individual and with that help them to reach their full potential. Several participants  identified that the concept is more complex than a metric or a protocol. For instance, during the interview, a participant emphasized that \emph{"all the time is spent on optimising against one of these metrics, leading to unfair outcomes being declared fair just due to the model performance on the metric"}.
Ideally, a common emerging view was that a system would be evaluated not just on the immediate intended effect (e..g, if an auto-generated hint helps students answer a question) but on broader educational goals (e.g., the student do well in classes). 

\subsection{Challenges and needs in team blind spots and practices (Q7)}

\vspace{2mm} \noindent \textbf{Support in the selection of the demographic groups to consider}. Several participants think that the decision on which demographic attributes to consider in an analysis is challenging. In line with this, while answering to question $7$ of the survey, a participant raised a doubt that \emph{"there might not be a need to add gender data to certain problems as it should not be relevant or, on the other hand, it should be included to show it is irrelevant"}. Another participant emphasized that \emph{"many studies do not consider social-economic characteristics of the students which may also bias the resulting models"}. In addition, interviews strengthened the view on the lack of datasets that do not just include all demographics, but also fair observations. For example, datasets containing salaries for people are probably still biased as salaries were decided by humans with their own biases. However, it is hard to define how an actual unbiased dataset should look like, and it is likely that all quantitative data have biases of some form or another. 
Overall, there is a need for standards on the demographic groups to consider. 

\vspace{2mm} \noindent \textbf{Building social and multi-disciplinary awareness in teams}. Several respondents highlighted that educational AI models are designed by technical people who often do not have the social science training to understand the socio-cultural implications of their algorithm designs. This might lead to prefer computational expediency over considerations of social justice. In this regard, a participant emphasized \emph{"the need for equity training and understanding for developers and researchers"}. 
Overall, our respondents call for inclusivity and diversity in the teams in charge of developing educational AI. In current practices, several participants identified that most of the AI models are constructed with a single mindset from an specific field (generally Computer Science). Since data and models are about and impact on people, it would be important to include other perspectives from social sciences to understand why and when the variables to be collected are valid. 

\section{Conclusion and Future Directions}

Though researchers and practitioners are already grappling with biases and unfairness in educational AI systems, research on this topic is rarely guided by a common understanding and view of the faced challenges and needs. In this work, we conducted the first systematic investigation of experts teams’ challenges and needs for support in developing fairer educational AI. Even when experts are motivated to improve fairness in their educational applications, they often face technical and organizational barriers. We highlight a few emerged aspects below:

\begin{itemize}[leftmargin=*]
\item Future research should also support experts in collecting and curating high-quality datasets, with an eye towards fairness in downstream AI models, reducing cultural dependencies in demographic representation. Moreover, large-scale data collection should be paired with an in-depth description of the local contexts, since biases are driven by complex reasons to be understood locally. Localized and causal data collection paired with data sharing practices are needed, posing attention in giving individuals control of their data.
\item Though fair educational AI has mainly focused on data collection, assessment and debiasing of unfairness is also an important area of work. Challenges and needs in this area include having continuous fairness assessment at all stages of the pipeline, understanding and acknowledging the potential weaknesses of the system, reducing frictions between model effectiveness and fairness, and creating cross-institutional frameworks for addressing fairness.
\item Domain-specific educational resources, metrics, processes, and tools are urgently needed. Challenges and needs in this perspective include, among others, practices for opening data and source code for public scrutiny, including fairness not only as a property of the AI model, showing explicit evidence the system’s potential unfair impacts, creating participatory processes for fairness checking, and defining responsibilities around fairness issues. 
\item The development of processes and tools for fairness-focused auditing is also important, to surface fairness issues in complex, multi-component educational AI systems. Among others, challenges and needs include fostering a more focused human-centered evaluation of fairness, contextualized and application-specific tools for auditing, and long-term learning-related auditing of fairness.
\item Finally, another area with several challenges and needs concern the teams working on educational AI. Among others, supporting the team in the selection of the demographic groups to consider and building multi-disciplinary awareness in teams are two of the more relevant aspects to work on. 
\end{itemize}

The rapidly growing area of fairness in educational AI presents many challenges and needs. The resulting systems are increasingly widespread, with proved potential to amplify social inequities, or even to create new ones. As research in this area progresses, it is urgent that research agendas are aligned with the challenges and needs of those who affect and are affected by educational AI systems. We view the directions outlined in this paper as critical opportunities for the AI and the educational research communities to play more active, collaborative roles in making real-world educational AI systems fair.

\vspace{2mm} \noindent \textbf{Acknowledgments}. Roberta Galici gratefully acknowledges the University of Cagliari for the financial support of her PhD scholarship.

\bibliographystyle{splncs04}
\bibliography{bib_file}

\end{document}